# LOW COST PAGE QUALITY FACTORS TO DETECT WEB SPAM

Ashish Chandra, Mohammad Suaib, and Dr. Rizwan Beg

Department of Computer Science & Engineering, Integral University, Lucknow, India

## ABSTRACT

*Web spam is a big challenge for quality of search engine results. It is very important for search engines to detect web spam accurately. In this paper we present 32 low cost quality factors to classify spam and ham pages on real time basis. These features can be divided in to three categories: (i) URL features, (ii) Content features, and (iii) Link features. We developed a classifier using Resilient Back-propagation learning algorithm of neural network and obtained good accuracy. This classifier can be applied to search engine results on real time because calculation of these features require very little CPU resources.*

## 1. INTRODUCTION

Internet has become one of the most important part of our life. Today we use Internet not only for entertainment but also for searching, banking, business, shopping and so on. But at the same time it has also imposed threats like spam, scam, malware, and phishing. Web spam is an economical phenomena where creation of spam is inversely proportional to cost of spam generation and distribution. Differentiating between desirable and undesirable content is a big challenge for users as well as for search engines. It is very important that we select features properly and carefully for web spam classification. According to researchers complex features like PageRank improves the classification performance marginally in machine learning process but it incurs high cost in terms of computing resources. We need more efficient, generic, and highly adaptive classifiers to detect web spam. The neural network based methods have high ability of generalization and adaption. In this paper we are presenting 32 low cost page quality factors which can be calculated in real time for a large number of web pages. We have discussed these factors in detail in section 3 of this paper. These factors can be used in any machine learning classification technique for efficiently detecting web spam. In this paper we have used Resilient Back-Propagation learning method of neural network to train our classifier

## 2. PREVIOUS WORK

Fetterly, *et al.*[1] used different content features like word count, title length, language model etc. They used in their experiment C4.5 Decision Tree, Bagging, and Boosting and obtained good results. They achieved true positive rate as 86.2% and true negative rate as 97.8% for a boosting of ten C4.5 decision trees.

Zhu and Wu [2] used reverse engineering to analyze SEO factors like page rank, URL, and Google search results. They highlighted mainly five factors which are: URL length, keywords in domain name, keyword density in H1, keyword density in title tag and keywords in URL.

Erdelyi *et al.* [3] used state of art learning models with LogitBoost and RandomForest. They used less computation hungry content features and investigated tradeoff between these features and spam detection accuracy. According to authors adding new features increases performance but complex features like PageRank improves classification performance marginally.

## 3. LOW COST PAGE QUALITY FACTORS

We are presenting 32 low cost page quality factors. These are categorized 3 categories: URL (10 features), Content (17 features) and Link (5 features). We call these features as low cost because these features require less computing resources in the extraction process. These features are listed below.

### 3.1. URL Features

#### 3.1.1. SSL Certificate

SSL refers to Secure Socket Layer. It is a standard for establishing an encrypted link between a web server and a web browser. It allows sensitive information to be transmitted securely. SSL Certificate ensures that the website is trusted and its owner has an identity. It incurs extra cost to the website owner. It makes spamming economically infeasible. The websites with SSL certificates use https protocol instead of plain http protocol.

#### 3.1.2. URL Length

Normally spam pages have very long URLs due to keyword stuffing in URL. So very long length URLs may represent probability of spam [1]. According to Zhu *et al.* [2] short length URLs are preferred by most search engines.

#### 3.1.3. URL Represents a Sub-Domain

Spammers create multiple sub-domains on a single domain to create multiple websites. By doing this they save on the cost of purchasing multiple domains and hosting charges. So the spam pages have higher probability that they are hosted on sub-domain.

#### 3.1.4. Authoritative TLD

There are certain Top Level Domains (TLD) such as .gov, .edu etc that can be registered only by legislatively recognized authorities. So if a TLD is authoritative then there is a very little chance that it is being used for spamming [1], [2].

#### 3.1.5. More than 2 Consecutive Same Letter in Domain

Normally spammers register domains in bulk using automated software. The websites hosted on them are for a short period of time and these websites are not for humans. These websites are created just for search engine crawlers to get high PageRank for the target page. These domain names contain random alphabets or digits. So if a domain name contains three or more same consecutive letters then it is a possibility that it is a spam. An example of such domain name can be cheaploanzzz.com.

### 3.1.6. More than Level 3 Sub-domain

The URL like 'microsoft.com.phishy.net' can be used for phishing or scam. A normal user may believe that this website is related to microsoft.com.

### 3.1.7. Many Digits or Special Symbols in Domain

If a domain name contains many digits and special characters then we can say that it is not a user friendly domain name. It may be created by automated software just to form link farms [1].

### 3.1.8. IP Address not Domain Name

Spammers  try to save cost by hosting website on bare IP address instead of purchasing domain name. IP address is not user friendly. Example of such URL can be 193.178.2.101/index.html.

### 3.1.9. Alexa Top 500 Website

Alexa is an organization which provides ranking of websites based on the traffic on them. If a website is on top 500 list then there is little chance that it has spam [4].

### 3.1.10. Domain Length

Normally spam websites have long domain name because of keyword stuffing [5]. Example of such website is http://www.buycheapextralongshowercurtains.com.

## 3.2. Content Features

### 3.2.1. HTML Length

Pages with large HTML are normally quality pages with rich content and good user experience. Such pages are normally not spam pages. Whereas pages with very little HTML size are generally spam [6].

### 3.2.2. Text Word Count

Normally pages with high word count up to a certain extent contain good information. Such pages have less probability of being spam [1]. But if the text word count is very high then again there is a possibility of spam

### 3.2.3. Text Character Length

High text character length up to a certain extent also represents content rich page. But very high text character length is a signal of spam.

### 3.2.4. Text to HTML Ratio

The lower percentage of visible text in the HTML of page suggests better quality of page due to better user experience.

### 3.2.5. Average Word Length

Spam pages have higher average word length due to the keyword stuffing. Spammers uses keywords extensively in place of pronouns, prepositions etc which increases average word length [1].

### 3.2.6. Existence of H2

H2 represents heading in large font size. Search engine gives preference to text with H2 considering it as keywords while indexing the page.

### 3.2.7. Existence of H1

H1 represents heading with very large text. Spammers heavily use it to lure users to perform an action. Search engines also give importance to it [2].

### 3.2.8. Video Integration

Video integration on a web page enhances users' experience. It is a quality factor for a web page. Normally good websites have embedded video from video sharing sites like YouTube, Vimeo etc.

### 3.2.9. Number of Ads

Spammers mostly create pages for money making so they put a number of advertisements on the page. Normally these pages contain thin content and are useless for a user [7], [8]. So large number of advertisement are signal of spam.

### 3.2.10. Title Length

Page with large title length may have keyword stuffing in title tag. Non-spam pages have title length within a limit [1], [9].

### 3.2.11. Compression Ratio of Text

If the compression ratio of visible text of the page is high (above 4) then it means page contains repeated keywords and phrases. It suggests that the page may be spam [1].

### 3.2.12. Use of Obfuscated Script

Obfuscated JavaScript code are used to hide pop-ups, redirections cloaking, and text hiding from search engine crawlers [10]. Example of such code is use of encode(), decode(), escape(), unescape() functions inside of one another [11]. According to Google's Search Engine Quality Engineer, Matt Cutts, [12] crawlers are not capable of fully executing JavaScript.

### 3.2.13. Description Length

The description meta tag provides brief information about content of the page to the search engines. Very small description or no description suggests that the page quality may not be good, whereas very large description suggests that it may contain keyword stuffing so there is chances of spam [6].

### 3.2.14. Image Count

The high quality pages normally have large number of images which leads to better user experience. Spam pages are generally created by automated software so it is unusual to have that number of images on such pages [6].

### 3.2.15. Presence of Alt Text for Image

The alt text is used to describe image where either image cannot be displayed on web browser or the page is being accessed by a blind person using an accessibility option. The presence of alt text for images on the page is a signal for better quality web page. Normally spammers do not put that effort on the page due to economical constraints. Other reason is that spammers generally generate these pages with automated software which do not go into this much details [6].

### 3.2.16. Call to Action

If a page contains high amount of call to action phrases (such as Act Now, Buy Now, Register Immediately, Limited Offer, Last Chance) then it is a signal of high monetization of the page or a page promoting scam or malware. These pages are designed to invoke a user to do certain action. This is a strong signal of page being a spam.

### 3.2.17. Stop Words

Normally spammers create pages with full of keyword stuffing. These pages have low percentage of stop words [6]. Stop words are that words which do not represent any useful meaning if used alone. Example of stop words are: it, this, being, have, he, now, such, is etc.

## 3.3. Link Features

### 3.3.1. Number of Internal Links

A good quality website has a good structure of internal links to connect other pages of the website so that a user can navigate easily [13].

### 3.3.2. Self Referential Internal Link

Spammers put multiple self referential links on same page with different keywords as anchor text to rank the same page for different keywords.

### 3.3.3. Number of External Links

Spammers use external links to authority pages to obtain high hub score by exploiting HITS algorithm [14].

### 3.3.4. Percentage of Anchor Text to Total Text

High percentage of anchor text to total text suggests that the page is just a part of link farm and has less information data [1].

### 3.3.5. Anchor Text Word Count

Higher average word count of anchor text suggests that there is a possibility of spamdexing with long keyword phrases. This is done because search engines give importance to anchor text while indexing a page for keywords [15].

## 4. EXPERIMENT AND RESULT

We used 32 page quality factors described in section 3 of this paper. We evaluated these factors by using a classifier based on Resilient Back-propagation Learning algorithm of Multilayer Perceptron Neural Network. The network we created has single hidden layer and the output layer with one neuron. Each neuron of the network uses bipolar sigmoid function with output range [-1 , 1].

$$f(x) \; = \; \frac{2}{1 + e^{-\alpha x}} \; - \; 1$$

The stopping criteria of training is number of iteration $\theta$ = 200

Number of neurons in hidden layer = 10

We tested performance of the classifier 20 times for each category and obtained average of each result category. The size of our dataset was 370 pages in which about 30% pages were spam and rest were ham pages. For the purpose of training of the classifier we randomly selected 300 pages from the dataset and for testing we selected remaining 70 pages.

We created following tables to show the performance result of each category of our quality factors.

Table 1. Performance Analysis.

| Features | Sensitivity | Specificity | Efficiency | Precision | F1Score | Accuracy |
|---|---|---|---|---|---|---|
| URL | 0.5051 | 0.9255 | 0.7153 | 0.8460 | 0.6272 | 0.7433 |
| Content | 0.7308 | 0.9000 | 0.8154 | 0.8489 | 0.7848 | 0.8267 |
| Link | 0.6461 | 0.9000 | 0.7731 | 0.8329 | 0.7266 | 0.7900 |
| URL+Content | 0.8615 | 0.9568 | 0.9092 | 0.9403 | 0.8986 | 0.9155 |
| URL+Link | 0.8410 | 0.9647 | 0.9028 | 0.9500 | 0.8906 | 0.9111 |
| Content+Link | 0.7423 | 0.8985 | 0.8204 | 0.8540 | 0.7917 | 0.8308 |
| URL+Content+Link | 0.8807 | 0.9529 | 0.9168 | 0.9357 | 0.9070 | 0.9216 |

The result given in Table 1 shows that when we used all the factors i.e. URL, Content, and Link features, we achieved best performance. The classification accuracy was 0.92, efficiency achieved was 0.91, precision was 0.93 and F1 score was 0.90.

## 5. CONCLUSION

In the experiment we can conclude that performance of the classifier improves with increased number of classification features. We have described an approach for classifying pages automatically as spam or ham based on supervised learning across URL, content, and link features. Our back-propagation learning neural network based classifier performed well and produced good result. The 32 low cost page quality factors can be used to detect web spam efficiently, accurately and economically. Since neural network learning is very adaptive and it performs well with noise as well, we can say that it can provide effective classification even in rapidly changing scenario of adversarial information retrieval like search engine results.